\documentclass[11pt,twoside]{article}
\usepackage{newpasp}
\usepackage{epsf}

\index{Stocke, J. T.}
\index{Shull, J. M.}
\index{Penton, S. V.}
\index{Gibson, B. K.}
\index{Giroux, M. L.}
\index{McLin, K. M.}

\begin{document}

\title{The Local Ly$\alpha$ Forest: H~I in Nearby Intergalactic Space}
\author{John T. Stocke, J. Michael Shull, Steven V. Penton, 
Brad K. Gibson,
Mark L. Giroux, Kevin M. McLin}
\affil{Center for Astrophysics \& Space Astronomy, and Dept. of Astrophysical
\& Planetary Sciences, University of Colorado, Boulder, CO 80309-0389}

\begin{abstract}
Detecting H~I using redshifted Ly$\alpha$ absorption lines is
$\sim$10$^6$ times more sensitive than using the 21\,cm emission line. 
We review recent discoveries of H~I Ly$\alpha$ absorbers made with the 
{\it Hubble Space Telescope} (HST)  which have allowed us a first 
glimpse at gas in local intergalactic space between us and the ``Great Wall''. 
Despite its mere 2.4\,m aperture, HST can detect absorbers with column 
densities as low as those found using Keck at high-$z$
(N$_{\rm HI} \approx 10^{12.5}$ cm$^{-2}$). New results that will be
discussed include: the evolution of absorbers with redshift, the location
of absorbers relative to galaxies (including the two-point correlation
function for absorbers), the metallicity of absorbers far from galaxies,
and the discovery of hot $\sim$10$^{5-6}$\,K (shock-heated?)  absorbers.
The unique ability of VLA H~I observations in discovering the nearest
galaxies to these absorbers is stressed. 
\end{abstract}


\section{Introduction}

Unlike virtually all other astronomical objects, Ly$\alpha$ absorbing
``clouds'' were first discovered at great distances ($z\geq$2) due to
cosmological redshifts and the near-UV atmospheric ``cutoff''. It has
only been with the advent of the {\it Hubble Space Telescope} (HST),
with access to the ultraviolet, that nearby
examples have been found and studied. One of the major unexpected
discoveries made during the first year of HST was that the numerous
Ly$\alpha$ absorption lines found at high-$z$ persist, albeit with fewer
numbers, into the present epoch (Bahcall et~al. 1991, 1993; Morris et~al. 1991; 
and subsequent HST QSO Absorption-Line Key Project papers by Jannuzi et~al.
1998 and Weymann et~al. 1998). 
Extrapolations of the steep redshift dependence of the number evolution of
the lines seen at early times had predicted that very few low-$z$
Ly$\alpha$ lines would be found.  While these absorbers likely account for
the majority of all baryons at $z\geq$2, their still substantial numbers
at $z$$\sim$0 imply that $\geq$20\% of all baryons remain in these clouds
locally (Shull, Penton \& Stocke 1999a; Penton et~al. 2000b;  
Dav\'e et~al. 1998, 2000). Thus,
any account of the present-day distribution of baryons must include an
accurate census of these clouds, and the mass associated with them, as
inferred from their column densities and physical extent.  Further, the
evolving thermodynamic properties of the intergalactic medium (IGM)  are
probed using these clouds, as measurements of their doppler widths ($b$)
reveal information about their temperatures.  Observations of the changing
$b$ distribution and metal content of Ly$\alpha$ absorbers with time
reveal the history of energy and metal injection into the IGM from
$z\approx6$ to the present. 

Although it is already clear that much or even most of the baryons could
be in the local IGM, this number is quite uncertain, depending as it does
on two poorly known quantities:  (1) {\bf cloud extent}: Physical
extent measurements come only from common absorbers found in close pairs
of QSO sightlines and suggest extents of 100-300
$h^{-1}$~kpc, but are difficult to interpret; see Impey (1999). 
Very large and elongated ``clouds''
are suggested by cosmological simulations (Cen et~al. 1994;  Dav\'e
et~al. 1999), but sightline pairs alone cannot measure ``cloud'' shapes; 
(2) {\bf ionized fraction}: Photoionized models of
Ly$\alpha$ absorbers use the local AGN luminosity function (e.g., Shull et~al. 1999)
or limits on IGM H~I cloud H$\alpha$ emission (e.g., Donahue et~al. 1995) 
to estimate the intensity and spectrum of the local IGM ionizing flux.
As has been stressed by several authors (Cen \& Ostriker 1999; Dav\'e
et~al. 2000), as many as 30-40\% of all baryons could be in a hot,
shock-ionized phase, outside galaxy clusters and groups.  If so, the
ionized fraction of many Ly$\alpha$ absorbers has been underestimated by
using photoionization models, and the majority of local baryons are in the
IGM, not in galaxies. However, one must take care not to ``double count''
individual absorbers that are detected both in Ly$\alpha$ and O~VI, but only
to make sure that an appropriate ionized fraction is used in the accounting. 

While the above census is ample reason for studying the local Ly$\alpha$
forest in detail, it is also only at low-$z$ that Ly$\alpha$ absorber
locations can be compared accurately with galaxy locations, so that the
relationship between these ``clouds'' and galaxies can be determined. The
degree to which absorbers correlate with galaxies
has been controversial; Lanzetta et~al (1995) and Chen
et~al. (1998) argue that the absorbers are the very extended halos of
galaxies (see also Lin et al. 2000; Linder 2000), 
while Morris et~al. 1993; Stocke et~al. 1995; Impey, Petry \&
Flint 1999; Dav\'e et~al. 1999 and Penton et~al. 2000b argue that the
absorbers are related to galaxies only through their common association
with large-scale gaseous filaments.

\section{Results}

Surprisingly, and luckily for local IGM research, the modest HST aperture
is competitive with the 10\,m aperture of the Keck Telescope ($+$HIRES
spectrograph) in detecting Ly$\alpha$ absorbers because much brighter
targets can be observed. Figure~1 shows an HST$+$STIS (Space Telescope
Imaging Spectrograph) far-UV spectrum of the bright BL Lac Object PKS
2005-489, which detects Ly$\alpha$ absorbers with column density, 
N$_{\rm HI} \geq 10^{12.5}$ cm$^{-2}$, as low as the best Keck HIRES data 
(e.g., Cowie et~al. 1995). For reference to other H~I work in this 
conference, these absorbers have $\sim$10$^6$ times lower column densities 
than the weakest detections
using the 21\,cm emission line. The results reported here come chiefly from an
on-going survey of the local Ly$\alpha$ ``forest'', which utlilizes spectra
like that shown in Figure~1, and which is being conducted by our group at
Colorado, in collaboration with J. van Gorkom (Columbia), C. Carilli
and J. Hibbard (NRAO), and R. Weymann and M. Rauch (OCIW). These
UV spectra also allow important studies (i.e., metallicities) of high
velocity clouds (HVCs) not possible by other methods and which bear critically
on the nature of these HVCs (see Gibson et al. contribution to this volume).

\begin{figure}
\plotfiddle{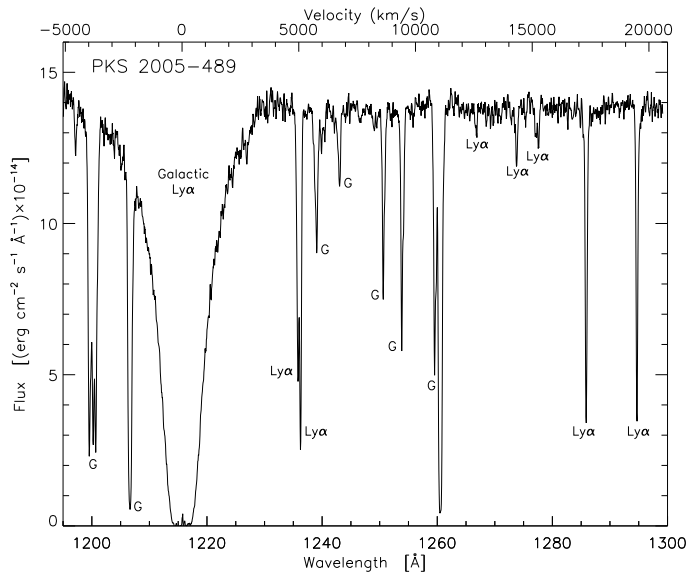}{3.0in}{0}{40}{40}{-128}{-5}
\caption{\small
An HST/STIS medium resolution (19 km s$^{-1}$) spectrum (Penton et~al. 2001,
in preparation) of the bright BL Lac Object PKS 2005-489 illustrates the best 
data obtained for this project. The deep, broad absorption at left center 
is the damped Ly$\alpha$ absorption due to the Milky Way. Other Galactic metal 
lines (S~II, Si~II, N~I, N~V, and Si~III) are marked with a ``G''. The
weakest Ly$\alpha$ absorbers have column densities comparable to the weakest
absorbers found in the best Keck spectra of high-$z$ QSOs,
N$_{\rm HI} = 10^{12.5}$ cm$^{-2}$. The heliocentric velocity scale at top is
for the Ly$\alpha$ absorbers only.}
\end{figure}

All specific results reported here use only the Goddard High Resolution
Spectrograph (GHRS) portion of our dataset, which includes 15 targets
(Penton et~al. 2000a,b, 2001).  Fifteen additional targets have now
been observed using STIS, which will extend the current results
significantly. From our GHRS survey the following results have been
obtained: 

\begin{enumerate}
\item Although only 116,000\,km\,s$^{-1}$ in pathlength has been observed
in our GHRS survey, we have detected 81 ($\geq$4$\sigma$) absorbers at
$cz\leq$20,000 km\,s$^{-1}$, yielding a
$dN/dz\sim$200 per unit redshift at N$_{\rm HI}\geq$10$^{13}$\,cm$^{-2}$ or 
one ``cloud'' every 20\,$h^{-1}_{75}$ Mpc! The 20\% baryon fraction quoted
above uses this line density, a 100 $h^{-1}_{75}$ kpc spherical cloud
extent and the standard 10$^{-23}$\,ergs\,s$^{-1}$\,
cm$^{-2}$\,Hz$^{-1}$\,
sr$^{-1}$ local
ionizing flux value (Shull et~al. 1999). Figure~2 compares line densities
as a function of redshift for two column density regimes: (1) 
N$_{\rm HI}\geq$10$^{14}$\,cm$^{-2}$ from the Bechtold (1994) 
high-$z$ compilation 
and the HST Key Project low-$z$ results of Weymann et~al. (1998); and (2)
10$^{13.1}\leq$N$_{\rm HI}\leq$10$^{14}$\,cm$^{-2}$ from Kim et~al. (1997) at
high-$z$ and our GHRS survey at $z\leq$0.067. The overall trend in these two
different column density regimes appears similar, and has been interpreted
using N-body + hydrodynamic simulations (Dav\'e et~al. 1999) as follows: 
(1) At $z\geq$1.6, the expansion of the Universe plus a nearly 
constant ionizing flux
greatly diminishes recombinations (i.e., the clouds are more highly
ionized at lower redshift to $z\approx$2), but 
(2) at $z\leq$1.6, the rapidly falling ionizing flux (due
to the diminished luminosities of AGN at $z\leq$1.6) partially offsets the 
decreasing recombinations, slowing the rapid evolution in $dN/dz$ seen at higher
$z$. But, there is some indication of added complications in that the two
column density regimes may evolve at different rates in $dN/dz$. At high-$z$,
the Kim et~al. data may have a shallower slope than the Bechtold data. 
At low-$z$, Weymann et~al. find that their lower column density
absorbers have a shallower slope in $dN/dz$ than their higher-N$_{\rm HI}$
absorbers. Spectra now being obtained by Jannuzi et~al. 
in HST Cycles 8 and 9 (PID \#8312 \& \#8673)
should address this issue by filling-in the missing data near the question
marks. It could even be that the lower column density
absorbers may actually increase in numbers between $z\sim$1.0 and 0. 

\begin{figure}
\plotfiddle{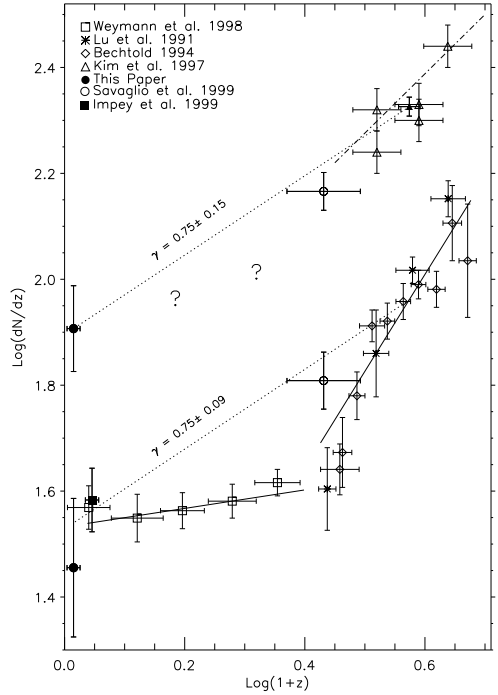}{3.5in}{0}{40}{40}{-120}{-35}
\caption{\small 
The Ly$\alpha$ absorber number density ($dN/dz$) as a function of redshift. The
overall trends for two different column density regimes 
(10$^{13.1}\leq$N$_{\rm HI}\leq$10$^{14}$\,cm$^{-2}$
at top and N$_{\rm HI}\geq$10$^{14}$\,cm$^{-2}$ at
bottom) are described in detail in the text.}
\end{figure}

\item The absorber $b$-value distribution at $z\sim$0 is similar
to that found at high-$z$, with a median $b$-value of 35\,km\,s$^{-1}$. No
obvious correlation between $b$ and N$_{\rm HI}$ is found. However,
when compared to $b$-values obtained from a curve-of-growth (COG) analysis using
higher-order Lyman lines 
(Shull et al.\ 2000) from spectra obtained with the Far UV Spectroscopic
Explorer (FUSE), Ly$\alpha$ line widths are a factor 
factor of two higher than the $b$-values inferred from the COG.
The FUSE data suggest that local
Ly$\alpha$ absorbers contain sizable nonthermal motions arising from
cosmological expansion and infall. Measuring the actual $b$-values of
these clouds is necessary to determine the IGM ``effective equation of
state'' (Hui \& Gnedin 1997; Ricotti, Gnedin \& Shull 2000) and thus when heat has
been input into the IGM. Higher resolution HST spectra and more FUSE
spectroscopy of low-column-density absorbers are now being obtained to
measure the $b$-values and clustering (see next item) of these clouds
more precisely. 
 
\item The two-point correlation function (TPCF), which measures
the clustering of Ly$\alpha$ absorbers, is similar to that found at
high-$z$ in that there is a 4$\sigma$ excess power over random at
$cz\leq$200 km s$^{-1}$. Impey, Petry \& Flint (1999) found a similar
result using lower resolution spectra. The absence of significant
clustering of these absorbers is strong evidence that Ly$\alpha$ clouds do
not arise in galaxy halos, although some investigators (Impey \&
Bothun 1997; Linder 2000) suggest that this may indicate that
low-surface-brightness (LSB) galaxy halos may be reponsible. 

\item Using available bright galaxy redshift surveys, we have
searched for the nearest known galaxies to these absorbers and have found
no close matches among a subset of 45 absorbers in sky regions surveyed
down to at least $L^*$.  Typical nearest-neighbor distances are several
hundred kpc to a few Mpc (median 1 Mpc) for 
H$_0$=75\,km\,s$^{-1}$\,Mpc$^{-1}$. Seven 
of these absorbers lie in well-defined galaxy voids, with no known
galaxies within 2-5\,$h^{-1}_{75}$\,Mpc. Deep optical and 21\,cm
observations still in progress have failed to locate any galaxies close to
these ``void'' absorbers (McLin et~al., this volume). 

\item The cumulative distribution of distances to nearest-neighbor galaxies
and the correlation of equivalent widths (EW) with impact parameters 
($\rho$) found by Penton et~al.\ (2001) are similar to those 
published previously (Stocke et~al.\ 1995). In the latter case, they
are similar to the results found by
Tripp, Lu \& Savage (1998) and Impey, Petry \& Flint (1999). The EW-$\rho$
correlation contains all the salient features (lack of correlation
at low-N$_{\rm HI}$) expected from the N-body+hydrodynamic simulations 
of Dav\'e et~al. (1999). Dav\'e et~al. interpret this plot as due to large-scale
structure filaments; the EW-$\rho$ correlation does {\bf not} 
require either a physical or a causal association 
with individual galaxy halos as proposed by Lanzetta et~al. (1995) and Lin
et~al. (2000). Our TPCF results and discovery of a substantial fraction 
($\sim$16\%) of
all absorbers in voids supports the Dav\'e et~al. (1999) interpretation. 

\item In one case, the sightline pair 3C273/Q1230$+$011 separated by
0.91$^\circ$ on the sky, we have a preliminary indication that both the 7
absorbers and 9 known galaxies in this vicinity are aligned along a
single ($>500\, h^{-1}_{75}$\,kpc), elongated ($>$3:1) filament at
$cz$=1000$-$2000\,km\,s$^{-1}$. This preliminary result (Penton et~al. 2001)
suggests that eventually, perhaps with the Cosmic Origins Spectrograph (COS)
on HST, we will be able to use Ly$\alpha$ absorbers and galaxy survey data
(e.g. Sloan Survey) to map out the full extent of large-scale structure
filaments in the local Universe. 

\item For one case, a close grouping of Ly$\alpha$ absorbers at  $cz\approx$
17,000\,km\,s$^{-1}$ in the direction of PKS 2155-304, we have good 
metallicity limits (Shull et~al.\ 1998) for low column density absorbers 
that lie far from galaxies. Figure 3 shows the PKS 2155-304 field,
overlaid with 21\,cm emission contours (also at $cz\sim$17,000\,km\,s$^{-1}$) 
from the VLA.  Deep optical galaxy survey work (McLin et~al.\ this volume) has
failed to find fainter galaxies closer to the absorbers than the H~I
emitters, reinforcing the importance of the VLA in this program.  No metal
lines (C~IV and Si~III)  have been detected as yet in the several strong
Ly$\alpha$ systems at this redshift, placing upper limits on the
metallicity of these clouds of $\la$1\% solar.

\begin{figure}
\plotfiddle{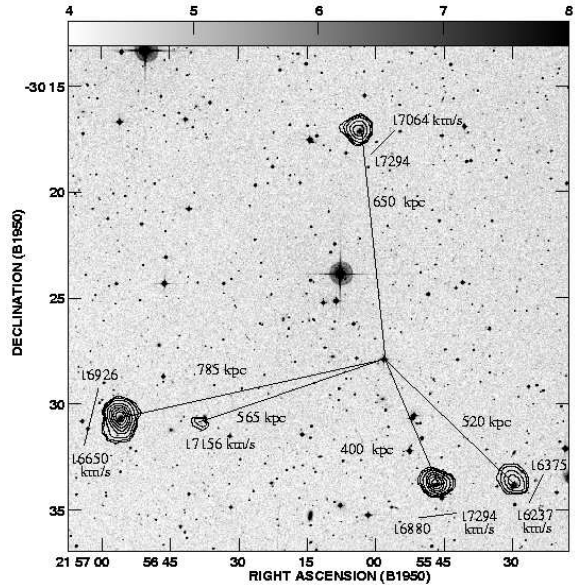}{3.0in}{0}{40}{40}{-110}{-50}
\caption{\small
An optical sky survey image with 21\,cm contours overlaid (Shull et~al.\
1998) shows the location of all H~I-emitting galaxies at
$cz\sim$17,000\,km\,s$^{-1}$. The lowest H~I
contours are at $1.65\times 10^{19}$\,cm$^{-2}$ and
total H~I masses as low as $0.8\times 10^9$\,M$_\odot$ are
detected. The central ``hub'' is the location of PKS 2155-304, while the
numbers on the ``spokes'' indicate the impact parameters
for each H~I galaxy assuming H$_0$=75 km s$^{-1}$ Mpc$^{-1}$.}
\end{figure}

The metallicity result in \#7 is still preliminary, awaiting better 
H~I column densities and metal-line measurements 
from new HST (STIS) and FUSE spectra. Also on-going is an
attempt to map the extent of metals in the IGM around galaxies using C~IV
observations for a subset of local, partially saturated 
absorbers found in our GHRS survey,
in conjunction with the available pencil-beam 
galaxy survey data. Preliminary results suggest that metals have been
spread up to $\sim$150\,$h^{-1}_{75}$\,kpc from galaxies (Stocke et~al.\
2001, in preparation), in agreement with simulations by Gnedin (1998). 
However, since we
have found at least one absorber that is undetected in C~IV, C~III and 
C~II but has strong O~VI absorption (Tripp et~al. 2001, in 
preparation), any 
metallicity result based on lower ionization species alone must be
viewed with caution. The existence of this one Ly$\alpha$ $+$ O VI absorber 
is additional evidence for these shock-heated ``clouds'' (see Tripp
contribution to this volume). We note that virtually all O~VI absorbers 
(the Dav\'e et~al. ``hot-warm'' phase) should be
detectable in Ly$\alpha$ and present in our GHRS survey. 
Therefore, it is important not to ``double count'' absorbers when determining
the total baryon content of the local Ly$\alpha$ ``forest''.
\end{enumerate}

Finally, the PKS 2155-304 field in Figure 3 underscores the important
role to be played by the VLA in this work. Virtually all
of the very close ($\rho\leq$200h$^{-1}_{75}$) absorber-galaxy pairs to
date are H~I-discovered galaxies, including the closest impact
parameter in our GHRS sample (100\,$h^{-1}_{75}$ kpc in the MRK335 sightline; 
van Gorkom et al. 1996) as well as the close proximity of the 
Haynes-Giovanelli H I cloud to the 3C273 and Q1230+011
sightlines (Penton et~al. 2001). Only H~I 21\,cm observations are 
unbiased against the
discovery of LSB galaxies, which have been suggested to be responsible
for some local Ly$\alpha$ absorbers.

\section{The Future}

Over the next few years, we expect that our own work, as well as
that of others, will increase the accuracy of every result in items \#1-5
above, including a revised value for the local baryon content of the IGM which
is probably accurate to 50\%. The $dN/dz$ relationship will be known in detail 
for both the high and low column density absorbers.
Future work on items \#6 and \#7 will determine whether
Ly$\alpha$ absorbers arise in large-scale structures or in very
extended galaxy halos. If the
former (which we strongly suspect based upon our own
investigations) some local Ly$\alpha$ absorbers 
will be found which were never ``polluted'' with metals from the
galaxies which co-habit the universal filamentary structure with them. 

\acknowledgments

We acknowledge the financial support at the University of Colorado of
grants provided through HST GO programs \#6593, \#8182 and \#8125, and NASA
Theory Grant NAG5-7262. We wish to
thank our collaborators in this work:  Ray Weymann, Jacqueline van Gorkom,
Chris Carilli, John Hibbard, Michael Rauch, and Jason Tumlinson.

\medskip
\noindent
{\bf Questions}

\smallskip
\noindent
{\it Linder:} 
We don't have a good sense of where all the galaxies are located
within the Dav\'e et~al. simulation (such as LSB galaxies). Furthermore, an
anti-correlation between equivalent width and impact parameter will be seen
whether absorbers arise in galaxies or not. Thus, the anti-correlation seen by
Dav\'e et~al. does not provide compelling evidence that absorbers do not arise
in galaxies.

\smallskip
\noindent
{\it Stocke:}
It is certainly true that the Dav\'e et~al. simulations locate galaxies
by a quite simple criterion that may have ``missed'' LSB galaxies.
However, it is striking that both the
observations and the simulations show similar slopes in the EW vs. $\rho$
relation as well as similar spreads in the data at both high and low column
density. You may need to explain why
16\% of Ly$\alpha$ absorbers are found in galaxy voids. We have looked
for faint galaxies near these ``void'' absorbers and not found any.
Also, to date, I know of no close association between any low-z
Ly$\alpha$ forest cloud and any LSB galaxy, despite sensitive attempts
to find them (e.g., Rauch, Morris \& Weymann; Impey, Petry \& Flint
1999; and our own H~I work in collaboration with van Gorkom and
Carilli). Because LSB galaxies are fairly abundant in H~I, our H~I surveys
near Ly$\alpha$ absorbers should have found some. So far we have not.

\medskip  
\noindent
{\it Meiksin:} 
The flattening in $dN/dz$ for the Ly$\alpha$ forest toward low
redshift can be accounted for predominantly as an ionization effect: the 
proximity effect tells us the ionizing background has decreased by an 
order of magnitude or more from $z=3$ to $z=0$. 
This decrease is independent of the nature of the
ionizing photon sources. 

\smallskip
\noindent
{\it Stocke:} I do not take the proximity effect results at 
$z \approx 0$ as definitive, but you are quite correct that the 
observational constraints (proximity effect at high $z$ and limits on 
H$\alpha$ emission from intergalactic H~I clouds at $z < 1$) 
argues for a substantial decrease in ionizing flux, regardless of 
the sources.

\medskip
\noindent
{\it Bland-Hawthorn:}
What about the possibility of condensation trails as the galaxies move through
their environment? You could imagine metal enriched material a megaparsec away
from the source.

\smallskip
\noindent
{\bf Stocke:} Yes, one could imagine such a thing. But the trick would be
to prove that this is what is going on. Indeed,
simulations (e.g., Gnedin 1998) show that mergers and winds can move
supernovae-enriched gas 100$-$200$h^{-1}$\,kpc away from their creation site and
cluster gas extends several hundreds of kpc from the cluster center.
Sensitive searches for metal enriched gas far away from galaxies are
difficult and have just begun, so until we find such gas, I am not too
concerned about explaining its precise origins.

\medskip
\noindent
{\it Katz:}
In the simulation, the Ly$\alpha$ absorbers come not from galaxies but from
structures containing galaxies. The simulations do include supernovae
feedback, but it doesn't produce winds.

\smallskip
\noindent
{\it Stocke:} Thanks, Neal, for that clarifiation. I add only that other
simulations (e.g., Gnedin 1998) do show that supernovae winds and 
galaxy merger events can move gas only about 100$-$200$h^{-1}$\,kpc
from galaxies, but nowhere near as far as the distances
we observe Ly$\alpha$ clouds to be due to those galaxies 
(e.g., the ``void'' absorbers).

\medskip
\noindent
{\it Disney:} 
Is the change in the ionizing flux used to explain the change of 
$dN/dz$ with $z$ a simulation result, or an observation? How secure is it?

\smallskip
\noindent
{\it Stocke:} Please see my reponse to Avery Meiksin's question. 
The proximity effect is a measured quantity, which can be used to
infer the mean ionizing background.  The error bars are significant
($\sim 50$\%), despite large samples of QSO absorbers (Scott et al.\ 2000). 
Other estimates of the ionizing intensity, $J_0$ at low redshift
are more indirect (Shull et al.\ 1999), and depend both on QSO
luminosity functions and radiative-transfer simulations.  
I would claim that the nearly two orders of magnitude drop in the ionizing
flux from $z \approx 2$ to $z = 0$ is quite secure observationally, 
although its mean value in the IGM, to say nothing of its dispersion, 
has not been measured at $z\approx$0 directly.

\end{document}